\documentstyle[12pt,fleqn]{article}

\def\eps{\varepsilon}
\def\eft{\tilde{E}_{\rm f}}
\def\ef{E_{\rm f}}
\def\tk{T_{\rm K}}
\def\vt{\tilde{V}}
\def\beeq{\begin{equation}}
\def\eneq{\end{equation}}
\def\beeqa{\begin{eqnarray}}
\def\eneqa{\end{eqnarray}}

\def\eps{\varepsilon}

\addtocounter{section}{-1}
\begin{document}

\begin{center}

{\large {\bf Theory of an elastic anomaly\\
in the heavy fermion system CeTe}\\
}

\vspace{1cm}

{\rm Kikuo Harigaya$^*$ and G. A. Gehring
}\\

\vspace{1cm}

{\sl Department of Physics, University of Sheffield,\\
Sheffield S3 7RH, United Kingdom}\\

\vspace{1cm}

(Received~~~~~~~~~~~~~~~~~~~~~~~~~~~~~~~~~~~)
\end{center}

\vspace{1cm}

\noindent
{\bf ABSTRACT}\\
The elastic anomaly observed in the coherent
Kondo state of Ce heavy fermion compounds
is analyzed by using the Anderson lattice model simulating
the energy level scheme of CeTe.  The $\Gamma_7$ doublets
and $\Gamma_8$ quartets of the 4f states are considered in the model.
We solve the mean field equations to derive the temperature
dependences of elastic constants, using the random phase
approximation like expression for the interaction between the
elastic strain and the crystalline field splitting.
We compare the calculation with the $(c_{11} - c_{12})/2$
and $c_{44}$ constants of CeTe.  The presence of the downward
dip and the observed overall temperature variations of the two
constants are well described by the present theory.  The origin
of the dip is the coupling between the elastic strain and the
splitting of the $\Gamma_8$ quartets.

{}~


\pagebreak


\section{INTRODUCTION}

Heavy fermion compounds show elastic anomalies in low temperatures.
They are related with crystalline field structures, magnetic phase
transitions, and so on.  The main contribution coming from the crystalline
fields is the peak or dip structures in the temperature dependences of
the elastic constants.  Most of the crystalline field splittings are
larger than the Kondo temperature and the anomalies occur in higher
temperatures than those of the coherent Kondo state.  For example,
the elastic constant $c_{33}$ of CeCu$_6$ [1] has a dip at about 10K.
This is due to the splitting larger than the Kondo temperature 4K.
However, a few compounds have the splittings which are comparable to
or smaller than the Kondo temperature.  The anomalies occur in the
coherent Kondo state.  The constants $(c_{11}-c_{12})/2$ and $c_{44}$
of CeTe [2] show the apparent dip at about 15K.  The $j = 5/2$
levels of Ce ions split into the $\Gamma_7$ Kramers doublet and the
$\Gamma_8$ quartet states.  The $\Gamma_7$ states are the ground states.
There is the splitting 30K between $\Gamma_7$ and $\Gamma_8$ states.
This is the origin of the dip.  Similarly, the $c_{44}$ constant of
the alloy Ce$_{0.5}$La$_{0.5}$B$_6$ shows the dip at about 0.2K, and
this temperature is lower than the Kondo temperature of this alloy [3].

The main purpose of the present paper is to develop a theoretical
description of the elastic anomaly by using a microscopic model.
We use the Anderson lattice model simulating the energy level
scheme of CeTe.  The $\Gamma_7$ doublets and $\Gamma_8$ quartets
are considered.  We solve the mean field equations and assume
that a strain field couples linearly
with the splitting of the $\Gamma_8$ states.
We calculate the linear susceptibility with respect to
the strain field.  Then, we derive the temperature
dependence of the elastic constant, using the random phase
approximation (RPA) like expression [4]
which includes the elastic susceptibility.  We compare the
calculation with the $(c_{11} - c_{12})/2$ and $c_{44}$ constants
of CeTe.  The observed overall temperature variations of the two
constants are well described by the present theory including the
coupling between the elastic strain and the splitting of the
$\Gamma_8$ quartets.

We explain the model in Section II.   We report the solution
in Section III and the elastic properties in Section IV.
We summarize the paper and give discussion in Section V.

\section{FORMALISM}

We formalize the infinite-$U$ Anderson lattice model
simulating the crystalline field structures of CeTe [2].
We use the slave boson method.  The model has the following form:
\beeqa
H &=& \sum_{i}
[ E_{\rm f} \sum_{l=1,2} + ( E_{\rm f} + \Delta - \delta ) \sum_{l=3,4}
+ ( E_{\rm f} + \Delta + \delta ) \sum_{l=5,6} ]
f_{i,l}^\dagger f_{i,l} \\ \nonumber
&+& \sum_{{\bf k},l=1-6} \eps_{\bf k} c_{{\bf k},l}^\dagger
c_{{\bf k},l} \\ \nonumber
&+& V \sum_{i,l=1-6} ( f_{i,l}^\dagger c_{i,l} b_i
+ b_i^\dagger c_{i,l}^\dagger f_{i,l} ) \\ \nonumber
&+& \sum_i \lambda_i ( \sum_{l=1-6} f_{i,l}^\dagger f_{i,l}
+ b_i^\dagger b_i - 1),
\eneqa
where $f_{i,l}$ is an annihilation operator of the f-electron of the
$l$-th orbital at the $i$-th site, $c_{{\bf k},l}$ is an operator of
the conduction electron with the wave number ${\bf k}$, and $b_i$ is
an operator of the slave boson which indicates the unoccupied state
at the f-orbital.  The atomic energy of the first and second orbitals
of f-electrons is $E_{\rm f}$; that of the third and fourth orbitals
is $E_{\rm f} + \Delta - \delta$; and that of the third and fourth
orbitals is $E_{\rm f} + \Delta + \delta$.  The two crystalline field
splitting parameters $\Delta$ and $\delta$ are considered in the model.
The first one $\Delta$ is the splitting between the ground state
$\Gamma_7$ doublet and the excited $\Gamma_8$ quartet states of
f-electrons.  The second splitting $\delta$ is due to the lattice
distortion from the cubic symmetry.  It is assumed that $\delta$
couples with a strain field $\eps$ linearly: $\delta = \eta \eps$,
where $\eta$ is the coupling constant.  For the conduction electrons,
the same quantum number is assumed as that of the f-electrons.  We
use the square density of states, $\rho \equiv 1/ND$, which extends
over the energy region, $-D < \eps_{\bf k} < (N-1) D$, where $N = 6$
is the total number of quantum states.  This assumes that the
combination $N \rho V^2$, which appears in the $1/N$ expansion, is
independent of $N$.  Therefore, the mean field theory becomes exact
as $N \rightarrow \infty$.  The third term in the hamiltonian is the
mixing interaction between f- and c-electrons, $V$ being the interaction
strength.  The last term limits the maximum number of f-electrons per
site up to unity.  This could be realized by the constraint
$\sum_{l=1-6} f_{i,l}^\dagger f_{i,l} + b_i^\dagger b_i = 1$ with the
Langrange multiplier field $\lambda_i$.

This model is treated within the mean field approximation:
$\langle b_i \rangle = r$, $\langle b_i^\dagger b_i \rangle
= r^2$, and $\lambda_i = \lambda $ (a site independent real value).
These mean field parameters are determined by solving the
following coupled equations [4]:\\
(1) the constraint condition,
\beeqa
& & \frac{1}{3D} \int dE
\frac{\vt^2}{(\eft + \Delta + \delta - E)^2} f(E-\mu)
\\ \nonumber
&+& \frac{1}{3D} \int dE
\frac{\vt^2}{(\eft + \Delta - \delta - E)^2} f(E-\mu)
\\ \nonumber
&+& \frac{1}{3D} \int dE
\frac{\vt^2}{(\eft - E)^2} f(E-\mu)
+ r^2 = 1,
\eneqa
(2) the self-consistency condition for $r$,
\beeqa
& & \frac{1}{3D} \int dE \frac{V^2}{E- \eft - \Delta - \delta} f(E- \mu)
\\ \nonumber
&+& \frac{1}{3D} \int dE \frac{V^2}{E- \eft - \Delta + \delta} f(E- \mu)
\\ \nonumber
&+& \frac{1}{3D} \int dE \frac{V^2}{E- \eft} f(E- \mu)
+ \lambda = 0,
\eneqa
and (3) the conservation condition of electron number $n_{\rm el}$,
\beeqa
& & \frac{1}{3D} \int dE
[1 + \frac{\vt^2}{(\eft + \Delta + \delta - E)^2}] f(E-\mu) \\ \nonumber
&+& \frac{1}{3D} \int dE
[1 + \frac{\vt^2}{(\eft + \Delta - \delta - E)^2}] f(E-\mu) \\ \nonumber
&+& \frac{1}{3D} \int dE
[1 + \frac{\vt^2}{(\eft - E)^2}] f(E-\mu)
= n_{\rm el},
\eneqa
where $f(x) = 1/[{\rm exp}(x/T) + 1]$ is the Fermi distribution
function, $\eft = \ef + \lambda$ is the effective f-level, and
$\vt = rV$ is the effective mixing interaction.  The integrations
are performed over all the energy region of the bands.  The three
equations are solved numerically for the three variables, $r, \lambda$,
and the Fermi level $\mu$.  In addition, the values at $T = 0$
can be obtained analytically.

\section{SOLUTION}

Equations (2), (3), and (4) are solved numerically for the parameters
$D = 5 \times 10^4$K, $V = 7500$K, $E_{\rm f} = - 10^4$K, and
$n_{\rm el} = 1.9$ as the typical values.  We take the splitting
parameter $\Delta=30$K. The close value for $\Delta$ has been
revealed in the experiment [5].
We consider the limit $\delta \rightarrow 0$ because the splitting
coming from the strain is so small and negligible.  As we will see
later, $\Delta$ is smaller than the Kondo temperature
$\tk = \eft - \mu$ at $T = 0$K.

Figure 1 shows the temperature dependences of parameters.
Figures 1 (a), (b), and (c) show the variations of $\eft$,
$\tk$, and the number of f-electrons per site $n_{\rm f}$,
respectively.  As the temperature increases, the order parameter
$r$ decreases, so that $n_{\rm f} = 1 - r^2$ increases.  The
quantity $r$ does not vanish even though the temperature is
much higher than $\tk$ (about 40K) at $T=0$.  This is the effect
of the change of the Fermi level $\mu$ to keep the total electron
number constant.  This effect has been reported previously [6,7].
According to the increase of $n_{\rm f}$, $\eft$ decreases,
which means the reduced itinerancy of f-electrons owing to the
increase of $n_{\rm f}$.  At low temperatures, the excitation
energy is limited by the smaller distance from the Fermi level
to the gap of the bands $l = 1,2$.  This results in the increased
value of $n_{\rm f}$ when the crystalline field is switched on.
Also, $\eft$ decreases and $\tk$ increases, due to the crystalline
field.  The similar dependence on $\Delta$ has been reported in the
previous paper [4].  The Kondo temperature $\tk (\Delta)$ (at $T=0$)
as a function of $\Delta$ satisfies the equation,
$[ \tk (\Delta) + \Delta]^2 \tk^2 (\Delta) = \tk^3 (0)$,
where $\tk (0) = D \exp [-D (\mu - E_{\rm f})/V^2]$ is the Kondo
temperature for $\Delta = 0$.  Starting from this analytic
expression, we could verify the low temperature variations of
parameters by using the expansion with respect to
$\Delta / \tk (0)$ assuming the small $\Delta$.

\section{ELASTIC ANOMALY IN LOW TEMPERATURES}

We shall discuss the change of elastic properties of heavy fermions
due to the crystalline field splitting in the low temperature
below $\tk$.  We shall calculate an elastic constant $c$
by the RPA-like formula [4] analogous to the plasmon excitation theory.
The constant $c$ is related with the linear susceptibility with respect
to $\delta$, as shown below:
\beeq
c = \frac{c_0}{1 + g \chi_\delta},
\eneq
where $c_0$ is the elastic constant of the system where there is
not interactions between the lattice and the electronic system,
and $g$ is the coupling constant.
There is a relation $g = c_0 \eta^2$, so $g$ is positive.
The analogous formula was used before [8] but in the linear response
theory.  We assume that $c_0$ is independent of the temperature.
The value of $g$ is unknown experimentally as well as theoretically.
In order to discuss the crystalline field effect on $c$, we treat
the factor $g$ as a kind of fitting parameters.  The quantity
$\chi_\delta$ is calculated as the second order derivative of
the mean field free energy:
\beeqa
\chi_\delta &=& - \frac{\partial^2 F}{\partial \delta^2}\\ \nonumber
&=& \frac{2}{3D} \int dE
\frac{\vt^2}{(\eft + \Delta - \delta - E)^3} f(E-\mu)
\\ \nonumber
&+& \frac{2}{3D} \int dE
\frac{\vt^2}{(\eft + \Delta + \delta - E)^3} f(E-\mu),
\eneqa
where the $\delta$ dependences of the band edges are neglected
in the derivatives because their effect is exponentially small.
In the actual calculation, we take the limit $\delta \rightarrow 0$,
because our problem is the elastic property at the equilibrium
of the cubic lattice where there is not the splitting $\delta$.

Figure 2 displays the temperature dependence of $\chi_0
\equiv {\rm lim}_{\delta \rightarrow 0} \chi_\delta$.
Figure 2(a) shows the
variation over wide temperatures, and Fig. 2(b) shows the
detailed structure in low temperatures.  There is a
peak around $T=15$K.  This is owing to the large degree of freedom
for electrons and the crystalline field $\Delta$.  The position
of the peak would depend on parameters, but here the position
agrees with that of the $(c_{11} - c_{12})/2$ constant of CeTe [2].
The appearance of the peak has been discussed in the previous
paper [4].  The susceptibility in high temperatures is nearly
inversely proportional to $T$, showing the Pauli paramagnetic behavior.
The value of $\chi_0$ at $T=0$ is analytically expressed as,
\beeq
\chi_0 = \frac{2 \tk (\Delta)}{[\tk (\Delta) + \Delta]
[3 \tk (\Delta) + \Delta ]},
\eneq
by using the Kondo temperature.

Now, we compare the calculated $c/c_0$ with the experiments.
We plot the temperature dependences $c/c_0$, which are obtained
from the experimental data of CeTe [2].  They are shown
by the dots.  Figures 3 (a) and (b) are for the
$(c_{11} - c_{12})/2$ and $c_{44}$ constants.  The experimental
$c_0$ depends on the temperature.  We used the linear dependence
used in Ref. 2 for the $(c_{11} - c_{12})/2$ mode. The quantity $c_0$ becomes
softer as the temperature increases.  However, we cannot use
the experimental $c_0$ for the $c_{44}$ mode [2] in order to
compare with the theory.  We rather use the increasing linear function:
$c_0 (T) = 0.696 + 9.89 \times 10^{-5} T (10^{11} {\rm erg}/{\rm cm}^3)$.
It seems strange that the constant becomes larger as $T$ rises.
But, this does not mean that the crystal becomes harder for
increasing $T$.  Several lattice constants can become harder
as $T$ increases. In fact, $c_{44}$ becomes harder, and
$c_{11}$ and $(c_{11} - c_{12})/2$ become softer, in CeTe.
In the two figures, the calculated $c/c_0$ is shown by the curve
for $g = 6.9$K.  The elastic constant decreases from much higher to lower
temperatures than $\tk$.  The decrease is almost proportional
to $1/T$.  The agreement is good enough.  The decrease is the effect
of the valence fluctuation.  There is a downward dip around 15-20 K
and the position agrees with the experiments, too.
The overall temperature dependences are well explained by
the same $g$ for the two constants.  Of course, both $c_0$
and $\eta$ are different for the two constants.  Thus,
this result should be regarded as a coincidence.

\section{SUMMARY AND DISCUSSION}

We have solved the mean field equations of the Anderson lattice
model with the crystalline field splitting between $\Gamma_7$ doublets
and $\Gamma_8$ quartets.  It has been assumed that the strain field
couples linearly with the splitting of the $\Gamma_8$ states.
We have calculated the linear susceptibility with respect to
the strain field.  Next, we have derived the temperature
dependence of the elastic constant, using the RPA-like expression
which includes the elastic susceptibility.  We have compared the
calculation with the $(c_{11} - c_{12})/2$ and $c_{44}$ constants
of CeTe.  The observed overall temperature variations of the two
constants are well described by the present theory including the
coupling between the elastic strain and the splitting of the
$\Gamma_8$ quartets.  We believe that the presence of the
peak in $\chi_0$ is not an artifact of the mean field theory.
In fact, the magnetic susceptibility of the exact solution
of the single site system has a peak when the number of the
degree of the freedom is larger than two.

In the actual compound, $V$ is anisotropic, i.e., it has an
angle dependence in the momentum space: $V = V(\theta, \phi)$.
The mean field equations, Eqs. (2), (3), and (4), change
only in one point: the angle average,
\beeq
\int \frac{d\theta d\phi}{4 \pi} V^2 (\theta,\phi),  \nonumber
\eneq
appears in the equations.  This angle average can be absorbed
in the present formalism by regarding the angle averaged $\sqrt{V^2}$
as the isotropic $V$ of Eq. (1).  Thus, the parameter value
$V$ used in this paper should be interpreted as an averaged one.

In Ref. 3, the alloy system Ce$_{0.5}$La$_{0.5}$B$_6$ has the downward
dip in the temperature dependence of the $c_{44}$ constant.  The variation
is very similar to that in Fig. 3.  The same mechanism of the elastic
anomaly discussed in this paper would work in this alloy system, too.
The elastic anomaly in magnetic alloys could be treated by a microscopic
theory by using the coherent potential approximation applied to the
Anderson alloy system [9]. This calculation will be an interesting
extension of this paper.

By applying the magnetic field, the atomic energy levels of Ce
will split further.  This will result in more structures in
temperature variations of the elastic constants.  Experimental
as well as theoretical information will be useful for
detailed understanding of the electronic properties of
heavy fermion systems.

{}~

\noindent
{\bf ACKNOWLEDGEMENTS}\\
The authors thank Professor T. Goto for sending them copies of
his papers and the fruitful discussion.  They also thank
Dr. H. Matsui for the useful discussion.  This work was done
while one of the authors (K.H.) was staying at the University
of Sheffield on leave from the Electrotechnical Laboratory, Japan.
He acknowledges the financial support for the stay obtained from
SERC and the hospitality that he obtained from the University.

\pagebreak

\begin{flushleft}
{\bf REFERENCES}
\end{flushleft}

\noindent
$*$ Permanent address: Fundamental Physics Section, Physical
Science Division, Electrotechnical Laboratory,
Umezono 1-1-4, Tsukuba, Ibaraki 305, Japan;
E-mail address: harigaya@etl.go.jp.\\
$[1]$ T. Goto, T. Suzuki, Y. Ohe, T. Fujimura, S. Sakutsume,
Y. Onuki, and T. Komatsubara, J. Phys. Soc. Jpn. {\bf 57}, 2612 (1988).\\
$[2]$ H. Matsui, T. Goto, A. Tamaki, T. Fujimura, T. Suzuki, and
T. Kasuya, J. Magn. Magn. Mater. {\bf 76/77}, 321 (1988).\\
$[3]$ S. Nakamura, T. Goto, H. Matsui, S. Sakatsume, and
S. Kunii, (preprint).\\
$[4]$ K. Harigaya and G. A. Gehring, J. Phys.: Condens. Matter
{\bf 57}, 5277 (1993).\\
$[5]$ J. Rossat-Mignod, J. M. Effantin, P. Burlet, T. Chattopadhyay,
L. P. Regnault, H. Bartholin, C. Vettier, O. Vogt, D. Ravot,
and J. C. Achart, J. Magn. Magn. Mat. {\bf 52}, 111 (1985).\\
$[6]$  S. M. M. Evans, T. Chung, and G. A. Gehring,
J. Phys.: Condens. Matter {\bf 1}, 10473 (1989).\\
$[7]$ K. Harigaya, {\sl Master Thesis}, University of Tokyo, (1988).\\
$[8]$ P. Levy, J. Phys. C: Solid State Phys. {\bf 6}, 3545 (1973).\\
$[9]$ K. Harigaya. J. Phys.: Condens. Matter {\bf 2}, 4623 (1990).\\

\pagebreak

\begin{flushleft}
{\bf FIGURE CAPTIONS}
\end{flushleft}

\noindent
FIG. 1.  Temperature dependences of the mean field solution:
(a) $\eft$, (b) $\tk$, and (c) $n_{\rm f}$.  Parameters are
$D = 5 \times 10^4$K, $V = 7500$K, $E_{\rm f} = - 10^4$K,
$n_{\rm el} = 1.9$, and $\Delta = 30$K.

{}~

\noindent
FIG. 2.  Temperature dependences of the linear susceptibility
$\chi_0$.  The parameters are the same as in Fig. 1.
The figure (b) shows the structures in low temperatures in (a).

{}~

\noindent
FIG. 3.  Temperature dependences of the elastic constant $c/c_0$
for $g = 6.9$K.  The parameters are the same as in Fig. 1.
In (a), the calculation is compared with the $(c_{11} - c_{12})/2$
constant of CeTe.  In (b), the comparison with the $c_{44}$ constant
is made. Experimental data are shown by the dots.

\end{document}